\documentclass[aps,prb,twocolumn,english,superscriptaddress,floatfix]{revtex4-1}
\usepackage[T1]{fontenc}
\usepackage{amsmath}
\usepackage{amssymb}
\usepackage{multirow}
\usepackage{graphicx}

\begin{document}

\title{Generalized magnetoelectronic circuit theory and spin relaxation at interfaces in magnetic multilayers}

\author{G. G. Baez Flores}

\author{Alexey A. Kovalev}

\affiliation{Department of Physics and Astronomy and Nebraska Center for Materials and Nanoscience, University of Nebraska-Lincoln, Lincoln, Nebraska 68588, USA}

\author{M. van Schilfgaarde}
\affiliation{Department of Physics, King's College London, Strand, London WC2R 2LS, United Kingdom}

\author{K. D. Belashchenko}
\affiliation{Department of Physics and Astronomy and Nebraska Center for Materials and Nanoscience, University of Nebraska-Lincoln, Lincoln, Nebraska 68588, USA}

\date{\today}

\begin{abstract}
Spin transport at metallic interfaces is an essential ingredient of various spintronic device concepts, such as giant magnetoresistance, spin-transfer torque, and spin pumping. Spin-orbit coupling plays an important role in many such devices. In particular, spin current is partially absorbed at the interface due to spin-orbit coupling. We develop a general magnetoelectronic circuit theory and generalize the concept of the spin mixing conductance, accounting for various mechanisms responsible for spin-flip scattering. For the special case when exchange interactions dominate, we give a simple expression for the spin mixing conductance in terms of the contributions responsible for spin relaxation (i.e., spin memory loss), spin torque, and spin precession. The spin-memory loss parameter $\delta$ is related to spin-flip transmission and reflection probabilities. There is no straightforward relation between spin torque and spin memory loss. We calculate the spin-flip scattering rates for N|N, F|N, F|F interfaces using the Landauer-B\"uttiker method within the linear muffin-tin orbital method and determine the values of $\delta$ using circuit theory.
\end{abstract}

\maketitle

\section{Introduction}

Spin-orbit coupling (SOC) plays an essential role at metallic interfaces, especially in the context of spin transport related phenomena such as giant magnetoresistance (GMR), \cite{Bass:JMMM2016,Bass.PrattJoPCM2007} spin injection and spin accumulation, \cite{Johnson.Silsbee:PRL1985} spin transfer torque, \cite{Ralph:JMMM2008} spin pumping, \cite{Mosendz.Pearson.eaPRL2010,Heinrich.Burrowes.eaPRL2011,Tserkovnyak.Brataas.ea:RMP2005} spin-orbit torque, \cite{Liu:Science2012,Demidov.Urazhdin.eaNM2012} spin Hall magnetoresistance (SMR), \cite{Chen.Takahashi.ea:2013} and spin Seebeck effect (SSE). \cite{Bauer.Saitoh.ea:NM2012,Adachi.Uchida.ea:RoPiP2013,Meier.Reinhardt.eaNC2015} The concept of the spin mixing conductance, originally introduced within the magnetoelectronic circuit theory, \cite{Brataas.Bauer.ea:PR2006} plays a very important role in describing the spin transport at magnetic interfaces. \cite{Weiler:Prl2013}

Nevertheless, the spin mixing conductance in its original form cannot account for various important contributions associated with spin-flip processes, \cite{Kovalev.Brataas.ea:2002,Haney.Lee.ea2013,Rojas-Sanchez.Reyren.eaPRL2014,Chen.ZhangPRL2015,Amin.Stiles2016,Amin.StilesPRB2016,Tao.Liu.eaSA2018} coupling to the lattice, \cite{Kovalev.Bauer.ea:2007,Haney.StilesPRL2010} and other effects associated with magnons. \cite{Arias.MillsPRB1999,Azevedo.Oliveira.eaPRB2000,Beens.Heremans.eaJoPDAP2018} One can generalize the concept of spin mixing conductance by considering spin pumping in the presence of spin-flip processes \cite{Tserkovnyak.OchoaPRB2017} or by considering the magnetoelectronic circuit theory in the presence of spin-flip scattering. \cite{Belashchenko.Kovalev.ea:PRL2016} So far such generalizations were not able to clarify the role of interfacial spin relaxation (usually referred to as spin memory loss or spin loss) in processes responsible for spin pumping and spin-transfer torque. Recent progress in first-principles calculations of interfacial spin loss \cite{Belashchenko.Kovalev.ea:PRL2016} suggests that an approach fully accounting for spin-nonconserving processes can be developed. Experimentally, a great deal of data is available on the relation between spin-orbit interactions and the efficiency of spin-orbit torque. \cite{PhysRevLett.116.126601,PhysRevB.92.064426,PhysRevLett.112.106602,PhysRevLett.122.077201} This data is often interpreted intuitively in terms of the spin memory loss parameter, \cite{Bass:JMMM2016} while lacking careful theoretical justification.

In this work, we develop the most general form of the magnetoelectronic circuit theory and apply it to studies of spin transport, concentrating on such phenomena as spin-orbit torque and interfacial spin relaxation in multilayers. We introduce a tensor form for the generalized spin mixing conductance describing spin-nonconserving processes, such as spin dephasing, spin memory loss, and spin precession. We numerically calculate parts of the spin mixing conductance responsible for the spin memory loss in N|N, F|N, F|F interfaces in the presence of spin-orbit interactions using the Landauer-B\"uttiker method based on linear muffin-tin orbital (LMTO) method. We show that the generalized spin mixing conductance can be also used to describe spin-orbit torque when exchange interactions dominate and the torque on the lattice can be disregarded.  Our results for the generalized spin mixing conductance suggest that two distinct combinations of scattering amplitudes are responsible for spin memory loss and torque, and in general there is no simple connection between the two.

The paper is organized as follows. In Sec. II, we develop a general formulation of the magnetoelectronic circuit theory in the presence of spin-flip scattering. In Sec. III, we apply the magnetoelectronic circuit theory to calculations of spin loss in (N$_{1}$N$_{2}$)$_\mathcal{N}$, (N$_{1}$F$_{2}$)$_\mathcal{N}$, or (F$_{1}$F$_{2}$)$_\mathcal{N}$ multilayers connected to ferromagnetic leads. In Sec. IV, we apply the magnetoelectronic circuit theory to spin-orbit torque calculations. Computational details are described in Sec. V, and the technicalities of the adiabatic embedding approach are detailed in Sec. VI. Section VII presents numerical results for the spin-flip transmission and reflection rates and area-resistance products for N|N, F|N, F|F interfaces. Section VIII concludes the paper.

\section{Generalized circuit theory}

\subsection{Formalism}

The magnetoelectronic circuit theory follows from the boundary conditions linking pairs of nodes in a circuit. \cite{Brataas.Bauer.ea:PR2006} Here we consider the general case, allowing spin-nonconserving scattering at interfaces between magnetic or non-magnetic metals due to the presence of spin-orbit interaction or non-uniform magnetization. The boundary condition at an interface between nodes 1 and 2, with arbitrary distribution functions $\hat{f}_{a}$ ($a=1,2$ labels the node), is:
\begin{equation}
\hat{I}_{2}=G_{0}\sum_{nm}\left[\hat{t}_{mn}^{\prime}\hat{f}_{1}(\hat{t}_{mn}^{\prime})^{\dagger}-\left(M_{2}\hat{f}_{2}-\hat{r}_{mn}\hat{f}_{2}(\hat{r}_{mn})^{\dagger}\right)\right],\label{eq:boundary_conditions}
\end{equation}
where $G_{0}=e^{2}/h$, $\hat{r}_{mn}$ is the spin-dependent reflection amplitude for electrons reflected
from channel $n$ into channel $m$ in node 2, $\hat{t}_{mn}^{\prime}$ is the spin-dependent transmission amplitude for electrons transmitted from channel $n$ in node 1 into channel $m$ in node 2, and the Hermitian conjugate is taken only in spin space. Equation~(\ref{eq:boundary_conditions}) can be easily rewritten for the current $\hat{I}_{1}$ in node $1$. For a ferromagnetic node, the spin accumulation is taken to be parallel to its magnetization. The matrices $\hat{r}_{mn}$ and $\hat{t}_{mn}^{\prime}$
are generally off-diagonal in spin space.

It is customary to assume that the distribution functions in the nodes, $\hat{f}_a=\hat{\sigma}^{0}f^{0}_a+\hat{\boldsymbol\sigma}\cdot\mathbf{f}^{s}_a$,
are isotropic, i.e., independent of $\mathbf{k}$. In this case Eq.\ (\ref{eq:boundary_conditions}) reduces to generalized Kirchhoff relations: \cite{Belashchenko.Kovalev.ea:PRL2016}
\begin{align}
I_{2}^{0} & = G^{cc}_2\Delta f^{0}+\mathbf{G}^{cs}_2\cdot\Delta\mathbf{f}^{s}-\mathbf{G}^{m}_2\cdot\mathbf{f}_{2}^{s},\label{eq:charge}\\
\mathbf{I}_{2}^{s} & =\mathbf{G}^{sc}_2\Delta f^{0}+\hat{{\cal G}}^{ss}_2\cdot\Delta\mathbf{f}^{s}-\hat{{\cal G}}^{m}_2\cdot\mathbf{f}_{2}^{s},\label{eq:spin}
\end{align}
where $\Delta f^{0}=f_{1}^{0}-f_{2}^{0}$ and $\Delta \mathbf{f}^{s}=\mathbf{f}_{1}^{s}-\mathbf{f}_{2}^{s}$
are interfacial drops of charge and spin components of the distribution
function, and $\hat{I_a}=(\hat{\sigma}^{0}I^{0}_a+{\hat{\boldsymbol{\sigma}}}\cdot\mathbf{I}^{s}_a)/2$. The conductances in Eq.\ (\ref{eq:charge}-\ref{eq:spin}) carry a subscript 2 emphasizing that they generally differ from their counterparts describing the currents in node 1; this subscript will be dropped where it doesn't lead to confusion. The conductances are related through $\mathbf{G}^{cs}=\mathbf{G}^{sc}-\mathbf{G}^{t}$, $\hat{{\cal G}}^{ss}=G^{cc}\hat\sigma^0-\hat{{\cal G}}^{t}$, $\mathbf{G}^{m}=\mathbf{G}^{t}+\mathbf{G}^{r}$, $\hat{{\cal G}}^{m}=\hat{{\cal G}}^{t}+\hat{{\cal G}}^{r}$ to the following scalar, vector, and tensor quantities:
\begin{align}
 & G^{cc}=2G_{0}\sum_{mn}\mathcal{T}_{mn}^{\nu\nu},\\
 & G_{i}^{t}=4G_{0}\sum_{mn}i\varepsilon_{ijk}\mathcal{T}_{mn}^{jk},\\
 & G_{i}^{r}=4G_{0}\sum_{mn}i\varepsilon_{ijk}\mathcal{R}_{mn}^{jk},\\
 & G_{i}^{sc}=2G_{0}\sum_{mn}(\mathcal{T}_{mn}^{i0}+\mathcal{T}_{mn}^{0i}+i\varepsilon_{ijk}\mathcal{T}_{mn}^{jk}),\\
 & \mathcal{G}_{ij}^{t}=2G_{0}\delta_{ij}^{kl}\sum_{mn}(\mathcal{T}_{mn}^{kl}+\mathcal{T}_{mn}^{lk}+i\varepsilon_{klp}[\mathcal{T}_{mn}^{0p}-\mathcal{T}_{mn}^{p0}]),\\
 & \mathcal{G}_{ij}^{r}=2G_{0}\delta_{ij}^{kl}\sum_{mn}(\mathcal{R}_{mn}^{kl}+\mathcal{R}_{mn}^{lk}+i\varepsilon_{klp}[\mathcal{R}_{mn}^{0p}-\mathcal{R}_{mn}^{p0}]),
\end{align}
where $\delta_{ij}^{kl}=\delta_{ij}\delta_{kl}-\delta_{ik}\delta_{jl}$, Latin indices $i,\dots,l$ denote Cartesian coordinates and $m$, $n$ the conduction channels, and repeated Cartesian indices are summed over here and below. In the above expressions, we defined the following combinations of scattering matrix elements: \begin{align}
&\mathcal{R}_{mn}^{\mu\nu}=\mbox{Tr}[(\hat{r}_{mn}\otimes\hat{r}_{mn}^{*})\cdot(\hat{\sigma}^{\mu}\otimes\hat{\sigma}^{\nu})]/4 ,\\
&\mathcal{T}_{mn}^{\mu\nu}=\mbox{Tr}[(\hat{t}'_{mn}\otimes\hat{t}_{mn}^{\prime*})\cdot(\hat{\sigma}^{\mu}\otimes\hat{\sigma}^{\nu})]/4,
\end{align}
where Greek indices can take values from 0 to 3.

In order to obtain the circuit theory equations (\ref{eq:charge}) and (\ref{eq:spin}) from Eq.~(\ref{eq:boundary_conditions}), we used the trace relations for Pauli matrices,  $\mbox{Tr}(\hat{\sigma}^{i}\hat{\sigma}^{j})=2\delta_{ij}$,
$\mbox{Tr}(\hat{\sigma}^{i}\hat{\sigma}^{j}\hat{\sigma}^{k})=2i\varepsilon_{ijk}$,
and $\mbox{Tr}(\hat{\sigma}^{i}\hat{\sigma}^{j}\hat{\sigma}^{k}\hat{\sigma}^{l})=2(\delta_{ij}\delta_{kl}+\delta_{il}\delta_{jk}-\delta_{ik}\delta_{jl})$.
The unitarity condition gives the following identities:
\begin{align}
&\sum_{mn}\hat{r}_{mn}\hat{r}_{mn}^{\dagger}+\hat{t}'_{mn}(\hat{t}'_{mn})^{\dagger}=M_{2}\hat{\sigma}^{0},\\
&\sum_{mn}\hat{r}'_{mn}(\hat{r}'_{mn})^{\dagger}+\hat{t}_{mn}(\hat{t}_{mn})^{\dagger}=M_{1}\hat{\sigma}^{0},\\
&\sum_{mn}\hat{r}_{mn}(\hat{r}_{mn})^{\dagger}+\hat{t}_{mn}(\hat{t}_{mn})^{\dagger}=M_{2}\hat{\sigma}^{0},\\
&\sum_{mn}\hat{r}'_{mn}(\hat{r}'_{mn})^{\dagger}+\hat{t}'_{mn}(\hat{t}'_{mn})^{\dagger}=M_{1}\hat{\sigma}^{0},
\end{align}
which relate the conductances defined for the two nodes separated by the interface as
$G^{cc}_1=G^{cc}_2$, $\mathbf{G}^{cs}_1=\mathbf{G}^{cs}_2+\mathbf{G}^{m}_2$, and $\mathbf{G}^{cs}_2=\mathbf{G}^{cs}_1+\mathbf{G}^{m}_1$.

The interface conductances in the magnetoelectronic circuit theory have to be renormalized by the Sharvin resistance for transparent Ohmic contacts \cite{PhysRevB.56.10805,PhysRevB.67.094421} which allows comparison between \emph{ab initio} studies and experiment. \cite{Kovalev.Bauer.eaPRB2006} The circuit theory in Eqs.~\eqref{eq:charge} and \eqref{eq:spin} can be generalized to account for the drift contributions in the nodes by renormalizing the conductances $G^{cc}$,  $\mathbf{G}^{cs}$, $\mathbf{G}^{sc}$, $\mathbf{G}^{m}$, $\hat{{\cal G}}^{ss}$, and $\hat{{\cal G}}^{m}$. This can be done by connecting nodes $1$ and $2$ to proper reservoirs with spin-dependent distribution functions $\hat{f}_{L}$ and $\hat{f}_{R}$ via transparent contacts. The currents in the nodes then become $\hat{I}_{1}=2G_0 \hat{M}_{1}(\hat{f}_{L}-\hat{f}_{1})$ and $\hat{I}_{2}=2G_0\hat{M}_{2}(\hat{f}_{2}-\hat{f}_{R})$, where $\hat{M}_{1(2)}$ describe the number of channels (in general spin-dependent) in the nodes. Effectively, this leads to substitutions $f_{1}^{\uparrow(\downarrow)} \rightarrow f_{1}^{\uparrow(\downarrow)}+I_1^{\uparrow(\downarrow)}/(2G_0 M_{1}^{\uparrow(\downarrow)})$
and $f_{2}^{\uparrow(\downarrow)} \rightarrow f_{2}^{\uparrow(\downarrow)}-I_2^{\uparrow(\downarrow)}/(2G_0 M_{2}^{\uparrow(\downarrow)})$ in Eqs.~\eqref{eq:charge} and \eqref{eq:spin}.

Finally, we note that the conductance $\hat{{\cal G}}^{m}$ describes various spin-nonconserving processes, such as spin dephasing, spin loss, and spin precession. Therefore, it can be interpreted as a tensor generalization of the spin mixing conductance \cite{Brataas.Nazarov.ea:PRL2000,Brataas:TEPJBMaCS2001,Brataas.Bauer.ea:PR2006} to systems with spin-flip scattering. In the limiting case described in Ref.\ \onlinecite{Tserkovnyak.OchoaPRB2017}, our definition reduces to the generalized tensor expression suggested there. However, our definition is more general as it can account for processes corresponding to spin precession and spin memory loss. Spin-nonconserving processes can also result in spin-charge conversion (i.e., spin galvanic effect), which is described by $\mathbf{G}^{m}$ and $\mathbf G^{cs}$ conductances. Furthermore, $\mathbf{G}^{sc}$ describes the conversion of charge imbalance into spin current (inverse spin galvanic effect), and  $\hat{\cal G}^{ss}$ is the tensor spin conductance.

\subsection{Spin-conserving F|N interface}

We now apply the generalized circuit theory to an F|N interface. In the special case of a spin-conserving interface, Eqs.~(\ref{eq:charge}) and (\ref{eq:spin}) should be invariant under $SO(3)$ rotations in spin space, which reproduces the spin-conserving circuit theory: \cite{Brataas.Nazarov.ea:PRL2000,Brataas:TEPJBMaCS2001,Brataas.Bauer.ea:PR2006}
\begin{align}
&\mathbf G^{m}=0, \\
&\mathbf G^{cs}=\mathbf G^{sc}= G^{sc} \mathbf{m},\\
&\hat{{\cal G}}^{ss}= G^{cc}\, \mathbf{m}\otimes\mathbf{m} ,\\
&\hat{{\cal G}}^{m}=2{\cal G}^{\uparrow\downarrow}_r(\hat{1}-\mathbf{m}\otimes\mathbf{m})+2{\cal G}^{\uparrow\downarrow}_i \mathbf{m} \times,\label{eq:mixing}
\end{align}
where the tensor $\mathbf{m}\otimes\mathbf{m}$ implements a projection onto the magnetization direction, and ${\cal G}^{\uparrow\downarrow}_r$ and ${\cal G}^{\uparrow\downarrow}_i$ are the real and imaginary parts of the spin-mixing conductance ${\cal G}^{\uparrow\downarrow}=G_0\sum_{mn}(\delta_{nm}-r_{mn}^{\uparrow\uparrow}r_{mn}^{\downarrow\downarrow*}-t_{mn}^{\uparrow\uparrow}t_{mn}^{\downarrow\downarrow*})$.

\subsection{General F|N interface}

To understand further the structure of current responses, we expand the vector and tensor conductances in powers of magnetization:
\begin{align}
&G_i^\alpha=G_i^{\alpha(0)}+G_{i,k}^{\alpha(1)}m_k+G_{i,kl}^{\alpha(2)}m_k m_l+\cdots, \\
&{\cal G}_{ij}^\beta={\cal G}_{ij}^{\beta(0)}+{\cal G}_{ij,k}^{\beta(1)}m_k+{\cal G}_{ij,kl}^{\beta(2)}m_k m_l+\cdots,
\end{align}
where $\alpha$ stands for $sc$, $cs$, $t$, $r$, or $m$, $\beta$ stands for $ss$, $t$, $r$, or $m$, and the tensors $G_i^{\alpha(0)}$, $G_{i,k}^{\alpha(1)}$, $G_{i,kl}^{\alpha(2)}$, ${\cal G}_{ij}^{\beta(0)}$, ${\cal G}_{ij,k}^{\beta(1)}$, ${\cal G}_{ij,kl}^{\beta(2)}$, etc.\ are invariant under the nonmagnetic point group of the system.

The circuit theory substantially simplifies for axially symmetric interfaces, which are common in polycrystalline heterostructures. Choosing the $z$ axis to be normal to the interface and applying the constraints corresponding to the $C_{\infty v}$ symmetry, we obtain the expansion of vector conductances $\mathbf{G}^{sc}$, $\mathbf{G}^{sc}$ and $\mathbf{G}^m$ to second order in $\mathbf{m}$:
\begin{align}
\vec G^\alpha =\begin{pmatrix}
m_x x_1^{\alpha(1)}+m_y m_z x_1^{\alpha(2)}\\ m_y x_1^{\alpha(1)} -m_x m_z x_1^{\alpha(2)}\\ m_z x_2^{\alpha(1)}
\end{pmatrix},\label{eq:Gv}
\end{align}
where $x_1^{\alpha(1)}$, $x_2^{\alpha(1)}$, and $x_1^{\alpha(2)}$ are arbitrary coefficients.
For the tensor conductances $\hat{{\cal G}}^{ss}$ and $\hat{{\cal G}}^m$ we obtain
\begin{align}\label{eq:Gt1}
&\hat{{\cal G}}^\beta=\begin{pmatrix}
x_1^{\beta(0)} & 0 & 0 \\
0 & x_1^{\beta(0)} & 0 \\
0 & 0 & x_2^{\beta(0)}
\end{pmatrix}\\ \label{eq:Gt2}
&+\begin{pmatrix}
0 & -m_z x_1^{\beta(1)} & m_y x_2^{\beta(1)} \\
m_z x_1^{\beta(1)} & 0 & -m_x x_2^{\beta(1)} \\
-m_y x_3^{\beta(1)} & m_x x_3^{\beta(1)} & 0
\end{pmatrix}\\
&+\begin{pmatrix}
m_x^2 x_1^{\beta(2)}+m_z^2 x_2^{\beta(2)} & m_x m_y x_1^{\beta(2)} & m_x m_z x_4^{\beta(2)} \\
m_x m_y x_1^{\beta(2)} & m_y^2 x_1^{\beta(2)}+m_z^2 x_2^{\beta(2)} & m_y m_z x_4^{\beta(2)} \\
m_x m_z x_5^{\beta(2)} & m_y m_z x_5^{\beta(2)} & m_z^2 x_3^{\beta(2)}\label{eq:Gt3}
\end{pmatrix}
\end{align}
where $x_1^{\beta(0)}$, $x_2^{\beta(0)}$, $x_1^{\beta(1)}$, $x_2^{\beta(1)}$, $x_3^{\beta(1)}$, $x_1^{\beta(2)}$, $x_2^{\beta(2)}$, $x_3^{\beta(2)}$, $x_4^{\beta(2)}$, and $x_5^{\beta(2)}$ are arbitrary coefficients.

The role of spin-flip scattering becomes the most transparent if both the magnetization and the spin accumulation are either parallel or perpendicular to the interface. In this case, the tensor and vector conductances in Eqs.~(\ref{eq:charge}) and (\ref{eq:spin}) can be simplified, and we arrive at the following relations for relevant components associated with the in-plane and perpendicular directions:
\begin{align}
  G^{cc}&=G_{0}(T_{\uparrow\uparrow}+T_{\downarrow\downarrow}+T_{\uparrow\downarrow}+T_{\downarrow\uparrow}),\label{cond1}\\
  G^{sc}&= G_{0}(T_{\uparrow\uparrow}-T_{\downarrow\downarrow}+T_{\uparrow\downarrow}-T_{\downarrow\uparrow}) ,\label{cond2}\\
  G^{t}&= 2G_{0}(T_{\uparrow\downarrow}-T_{\downarrow\uparrow}) ,\,
 G^{r}= 2G_{0}(R_{\uparrow\downarrow}-R_{\downarrow\uparrow}) ,\label{cond3}\\
  \mathcal{G}^{t}&=2G_{0}(T_{\uparrow\downarrow}+T_{\downarrow\uparrow}),\, \mathcal{G}^{r}=2G_{0}(R_{\uparrow\downarrow}+R_{\downarrow\uparrow}), \label{cond4}
\end{align}
along with $G^{cs}=G^{sc}-G^{t}$, $\mathcal{G}^{ss}=G^{cc}-\mathcal{G}^{t}$, $G^{m}=G^{t}+G^{r}$, and $\mathcal{G}^{m}=\mathcal{G}^{t}+\mathcal{G}^{r}$. Of course, all quantities in these expressions are different for the in-plane and perpendicular orientations of the magnetization; the corresponding index has been dropped to avoid clutter. The spin-resolved dimensionless transmittances and reflectances
\begin{align}
T_{\sigma\sigma'}&=\sum_{mn}t_{mn}^{\sigma\sigma'}(t_{mn}^{\sigma\sigma'})^*,\label{tss}\\ R_{\sigma\sigma'}&=\sum_{mn}r_{mn}^{\sigma\sigma'}(r_{mn}^{\sigma\sigma'})^*\label{rss}
\end{align}
are defined in the reference frame with the spin quantization axis aligned with the magnetization.

Eqs.~(\ref{cond1})-(\ref{cond4}), together with Eqs.~(\ref{eq:charge}) and (\ref{eq:spin}), are also valid for axially symmetric F|F interfaces, as long as the magnetizations of the two ferromagnets are collinear. These expressions generalize the result given in Ref.\ \onlinecite{Belashchenko.Kovalev.ea:PRL2016} for axially symmetric N|N junctions to include F|N and F|F interfaces.

\subsection{Relation to Valet-Fert theory}

The Valet-Fert model \cite{Valet.Fert:1993} incorporates spin relaxation in diffusive bulk regions but makes restrictive approximations for the interfaces, treating them as transparent, spin-conserving, and prohibiting transverse spin accumulation. \cite{Kovalev.Brataas.ea:2002,Eid.Portner.eaPRB2002,Barnas.Fert.eaPRB2005,Urazhdin.Loloee.eaPRB2005,Bass.PrattJoPCM2007,Liu.Yuan.eaPRL2014} When spin relaxation at interfaces is of interest, the treatment based on the Valet-Fert model is forced to replace the interfaces by fictitious bulk regions, \cite{Bass.PrattJoPCM2007,Bass:JMMM2016} which is restrictive even for N|N interfaces. \cite{Belashchenko.Kovalev.ea:PRL2016}

Here we show how diffusive bulk regions can be incorporated in the generalized circuit theory. By introducing nodes near the interfaces and treating both interfaces and bulk regions as junctions, the generalized Kirchhoff's rules \cite{Kovalev.Brataas.ea:2002,Eid.Portner.eaPRB2002,Barnas.Fert.eaPRB2005,Urazhdin.Loloee.eaPRB2005,Bass.PrattJoPCM2007,Liu.Yuan.eaPRL2014} can be used to analyze entire devices with spin relaxation in the diffusive bulk regions and arbitrary spin-nonconserving scattering at interfaces.

The Valet-Fert model employs the following equations to describe spin and charge diffusion in a normal metal:
\begin{align}
  \partial_{x}^{2}(Df_{0}^{N})&=0,\label{eq:diffusion-N1}\\
  \frac{\partial^{2}}{\partial x^{2}}(D\mathbf{f}_{s}^{N})&=\dfrac{\mathbf{f}_{s}^{N}}{\tau_{sf}^{N}},\label{eq:diffusion-N2}
\end{align}
and in a ferromagnet:
\begin{align}
  \frac{\partial^{2}}{\partial x^{2}}(D_{\uparrow}f_{\uparrow}+D_{\downarrow}f_{\downarrow})&=0,\label{eq:diffusion-F1}\\
  \frac{\partial^{2}}{\partial x^{2}}(D_{\uparrow}f_{\uparrow}-D_{\downarrow}f_{\downarrow})&=\frac{f_{\uparrow}-f_{\downarrow}}{\tau_{sf}^{F}}.\label{eq:diffusion-F2}
\end{align}
Here $\mathbf{f}_{s}^{F}=\mathbf{m}(f_{\uparrow}-f_{\downarrow})/2$ is the spin accumulation in the ferromagnet, and the spin-flip relaxation times $\tau_{sf}^{N}=(l_{sf}^{N})^{2}/D$
and $\tau_{sf}^{F}=(l_{sf}^{F})^{2}(1/D_{\uparrow}+1/D_{\downarrow})/2$
are given in terms of the spin-diffusion lengths $l_{sf}^{N}$, $l_{sf}^{F}$ and diffusion coefficients $D$, $D_\sigma$. We now consider three basic circuit elements.

\subsubsection{Diffusive N region}

For a diffusive N layer, the solution of Eqs. (\ref{eq:diffusion-N1}) and (\ref{eq:diffusion-N2}) leads to a simplified version of Eqs. (\ref{eq:charge}) and (\ref{eq:spin}) with vanishing
vector conductances $\mathbf{G}^{sc}$, $\mathbf{G}^{sc}$, $\mathbf{G}^{m}$, and all tensor conductances reduced to scalars:
\begin{align}
  G_{N}^{cc}&=\frac{2D}{t_{N}},\\
  {\cal G}_{N}^{ss}&=G_{N}^{cc}\,\frac{\delta_{N}}{\sinh\delta_{N}},\\
  {\cal G}_{N}^{m}&=G_{N}^{cc}\,\delta_{N}\tanh\frac{\delta_{N}}{2},
\end{align}
where $t_{N}$ is the thickness of the N layer, and $\delta_N=t_{N}/l_{sf}^{N}$.

\subsubsection{Diffusive F region}

For a diffusive F layer with spin accumulation that is parallel to the magnetization, the solution of Eqs.~(\ref{eq:diffusion-F1}) and (\ref{eq:diffusion-F2}) leads to vanishing
$\mathbf{G}^{m}$ and the other conductances defined as follows:
\begin{align}
  G_{F}^{cc}&=(D_{\uparrow}+D_{\downarrow})/t_{F},\label{eq:F1}\\
  \mathbf{G}_{F}^{sc}&=\mathbf{G}_{F}^{cs}=\mathbf{m}(D_{\uparrow}-D_{\downarrow})/t_{F},\label{eq:F2}\\
  {\cal G}_{F}^{ss}&=G^*_F\frac{\delta_{F}}{\sinh\delta_{F}}+\frac{(G^{sc}_F)^2}{G^{cc}_F},\label{eq:F3}\\
  {\cal G}_{F}^{m}&=G^*_F\,\delta_{F}\tanh\frac{\delta_{F}}{2},\label{eq:F4}
\end{align}
where $G^*_F=[(G^{cc}_F)^2-(G^{sc}_F)^2]/G^{cc}_F$ is the effective conductance and $t_{F}$ the thickness of the F layer, and $\delta_F=t_{F}/l_{sf}^{F}$.

\subsubsection{Diffusive F|N junction}

As a simple application, consider a composite junction consisting of F and N diffusive layers separated by a transparent interface. Such an idealized junction can be used to model an interface with spin-flip scattering between F and N layers. \cite{Kovalev.Brataas.ea:2002,Eid.Portner.eaPRB2002,Barnas.Fert.eaPRB2005,Urazhdin.Loloee.eaPRB2005,Bass.PrattJoPCM2007,Liu.Yuan.eaPRL2014} Combining the results for F and N regions with boundary conditions, we find $\mathbf{G}^{m}=0$ and the following effective conductances:
\begin{align}
 & G^{cc}=(1/G_{F}^{cc}+1/G_{N}^{cc})^{-1},\label{eq:FN1}\\
 & \mathbf G^{cs}=\mathbf G^{sc}= \mathbf{G}_{F}^{sc} ,\label{eq:FN2}\\
 & \hat{{\cal G}}^{ss}=\frac{{\cal G}_N^{ss} {\cal G}_F^{ss}}{{\cal G}_N^{ss}+{\cal G}_F^{ss}+{\cal G}_N^{m}+{\cal G}_F^{m}} \mathbf{m}\otimes\mathbf{m},\label{eq:FN3}\\
 & \hat{{\cal G}}^{m}={\cal G}_{N}^{ss}+{\cal G}_{N}^{{m}}-\dfrac{{\cal G}_N^{ss} ({\cal G}_N^{ss}+{\cal G}_F^{ss})}{{\cal G}_N^{ss}+{\cal G}_F^{ss}+{\cal G}_N^{m}+{\cal G}_F^{m}} \mathbf{m}\otimes\mathbf{m},\label{eq:FN4}
\end{align}
where the conductances for the F and N layers should be taken from the previous subsections. If spin-flip scattering is negligible, we recover the known result: \cite{Kovalev.Brataas.ea:2002} $\hat{{\cal G}}^{m}={\cal G}_{N}^{ss}(1-\mathbf{m}\otimes\mathbf{m})$.

\section{Spin loss at interfaces}

The experimental data on interfacial spin relaxation comes primarily from the measurements of magnetoresistance in (N$_{1}$N$_{2}$)$_\mathcal{N}$, (N$_{1}$F$_{2}$)$_\mathcal{N}$, or (F$_{1}$F$_{2}$)$_\mathcal{N}$ multilayers connected to ferromagnetic leads, \cite{Bass.PrattJoPCM2007,Bass:JMMM2016} where $\mathcal{N}$ is the number of repetitions. The results have been reported \cite{Bass.PrattJoPCM2007,Bass:JMMM2016} in terms of the effective \textit{spin memory loss parameter} $\delta_N$ or $\delta_F$ obtained by treating the interface as a fictitious bulk layer and fitting the data to the Valet-Fert model. Here we relate the experimentally measured parameter $\delta_N$ or $\delta_F$ to the generalized conductances appearing in Eqs.~(\ref{eq:charge}) and (\ref{eq:spin}). We assume that the interfaces are axially symmetric and that the magnetization and spin accumulation are either parallel or perpendicular to the interface.

\subsection{N|N multilayer}

We first consider a multilayer with repeated interfaces between normal metals N$_{1}$ and N$_{2}$. We would like to assess the decay of spin current which may include the spin relaxation both at interfaces and in the bulk. To this end, we place nodes in both N$_{1}$ and N$_{2}$ layers and consider the case of axially
symmetric interfaces corresponding to relations, $\mathbf{G}^{sc}=\mathbf{G}^{sc}=\mathbf{G}^{m}=0$.
The relevant conductances $G^{cc}$, $\mathcal{G}^{ss}$, $\mathcal{G}_{1}^{m}$, and $\mathcal{G}_{2}^{m}$
account for the scattering in the bulk and/or at the interfaces. Using Eq.~(\ref{eq:spin}), we arrive at the following equations for the spin current in some arbitrary node $i$ in the superlattice:
\begin{align}
I_{i}^{s} & =\mathcal{G}^{ss}(f_{i-1}^{s}-f_{i}^{s})-\mathcal{G}_{i}^{m}f_{i}^{s},\label{SL1}\\
I_{i}^{s} & =\mathcal{G}^{ss}(f_{i}^{s}-f_{i+1}^{s})+\mathcal{G}_{i}^{m}f_{i}^{s},\label{SL2}
\end{align}
which results in the recursive formula:
\begin{equation}
\frac{2\mathcal{G}_{i}^{m}}{\mathcal{G}^{ss}}f_{i}^{s}=f_{i-1}^{s}-2f_{i}^{s}+f_{i+1}^{s}.
\label{eq:rec_app}
\end{equation}
This equation has analytical solutions:
\begin{equation}
f_{s}^{i}=C_{1}e^{\delta i}+C_{2}e^{-\delta i},\label{eq:sol}
\end{equation}
where the constants $C_{1}$ and $C_{2}$ are determined by the boundary conditions.
In the limit of weak spin-flip scattering, we obtain the leading term for the decay rate:
\begin{equation}
\delta^{2}\approx\frac{\mathcal{G}_{1}^{m}+\mathcal{G}_{2}^{m}}{ G^{cc}},
\label{eq:approx}
\end{equation}
where the constants $C_{1}$ and $C_{2}$ are defined by the boundary conditions. Note that to the lowest order in the spin-flip processes, only denominator in Eq.~\eqref{eq:approx} needs to be renormalized  by the Sharvin resistance for transparent Ohmic contacts, i.e., $1/\tilde G^{cc}=1/G^{cc}-(1/M_1+1/M_2)/(4 G_0)$. It is clear that the constant $\delta$ describes how the spin current decays as we increase the number of layers in the superlattice. The conductances in Eq.~(\ref{eq:approx}) may also include scattering in the bulk where the total conductances can be calculated by concatenating the corresponding bulk and interface conductances using Eqs.~(\ref{eq:charge}) and (\ref{eq:spin}). When obtaining $\delta$ from experimental data, one typically considers only interfacial contributions in Eq.~\eqref{eq:approx}, while the bulk contributions are simply removed. \cite{Bass:JMMM2016} This does not cause any problem when spin-orbit interaction is weak as in this limit the total  $\mathcal{G}^{m}$ is a simple sum of contributions from interface and bulk.

\subsection{F|N and F|F multilayers}

By considering F|N and F|F multilayers connected to ferromagnetic leads one can also quantify spin relaxation at magnetic interfaces. \cite{Bass:JMMM2016} In this case, a parameter $\delta$ describing the decay of spin current can also be related to the scattering matrix elements and to the generalized conductances in Eq. (\ref{eq:charge}) and (\ref{eq:spin}). We assume that we have a superlattice with repeated interfaces between normal (N$_1$) and ferromagnetic (F$_2$) layers. Normal can be considered a special case of F in this section, equations derived below also apply to F|F multilayers without any modifications. We would like to assess the decay of spin current due to spin relaxation at interfaces and in the bulk. We take nodes in F and N layers and consider the case of axially symmetric interfaces. We also assume collinear spin transport with the magnetization being in-plane or perpendicular to interfaces. The generalized conductances may include scattering both in the bulk and at the interfaces. Using Eqs.~(\ref{eq:charge}) and (\ref{eq:spin}), we arrive at the following equations for the spin and charge currents in node $i$:
\begin{align}
I_{i}^{0} & =G^{cc}(f_{i-1}^{0}-f_{i}^{0})+G^{cs}_{i-1}(f_{i-1}^{s}-f_{i}^{s})-G_{i}^{m}f_{i}^{s},\label{SL1-1}\\
I_{i}^{0} & =G^{cc}(f_{i}^{0}-f_{i+1}^{0})+G^{cs}_{i+1}(f_{i}^{s}-f_{i+1}^{s})+G_{i}^{m}f_{i}^{s},\\
I_{i}^{s} & =G_{i-1}^{sc}(f_{i-1}^{0}-f_{i}^{0})+\mathcal{G}^{ss}(f_{i-1}^{s}-f_{i}^{s})-\mathcal{G}_{i}^{m}f_{i}^{s},\\
I_{i}^{s} & =G_{i+1}^{sc}(f_{i}^{0}-f_{i+1}^{0})+\mathcal{G}^{ss}(f_{i}^{s}-f_{i+1}^{s})+\mathcal{G}_{i}^{m}f_{i}^{s},\label{SL2-1}
\end{align}
which results in the recursive formula:
\begin{equation}
\frac{2\mathcal{G}_{i}^{m}/G_{i-1}^{sc}-2G_{i}^{m}/G^{cc}}{\mathcal{G}^{ss}/G_{i-1}^{sc}-G^{cs}_{i-1}/G^{cc}}f_{i}^{s}=f_{i-1}^{s}-2f_{i}^{s}+f_{i+1}^{s},\label{eq:rec_app-1}
\end{equation}
Similar to non-magnetic case, the above equation has analytical solutions:
\begin{equation}
f_{s}^{i}=C_{1}e^{\delta i}+C_{2}e^{-\delta i}.\label{eq:sol-1}
\end{equation}
In the limit of weak spin-flip scattering, we obtain the leading term for the
decay rate:
\begin{equation}
\delta^{2}\approx\frac{\mathcal{G}_{F}^{m}+\mathcal{G}_{N}^{m}}{G^*},
\label{eq:approx-1}
\end{equation}
where $G^*=[(G^{cc})^{2}-(G^{sc})^{2}]/G^{cc}$ is the effective conductance of the scattering region. Note that to the lowest order in the spin-flip processes, only denominator in Eq.~\eqref{eq:approx-1} needs to be renormalized  by the Sharvin resistance for transparent Ohmic contacts, i.e., $1/\tilde G^*=1/G^*-(1/M_1^\uparrow+1/M_1^\downarrow+1/M_2^\uparrow+1/M_2^\downarrow)/(8 G_0)$. The constant $\delta$ describes how the spin current decays as we increase the number of layers in the multilayers. The conductances in Eq.~(\ref{eq:approx-1}) may also include scattering
in the bulk. The bulk and interface conductances can be concatenated using Eqs.~(\ref{eq:charge}) and (\ref{eq:spin}).

\section{Spin-orbit torque}

The discontinuity of spin-current at the interface following from the circuit theory in Eqs.\ \eqref{eq:charge} and \eqref{eq:spin} can be used to calculate the total torque transferred to both the magnetization and the lattice. In general, separating these two contributions is not possible without considerations beyond the circuit theory.
When exchange interactions dominate and the torque on the lattice can be disregarded, we can use the circuit theory to calculate the spin torque on magnetization. Note that spin-flip scattering and spin memory loss can still be present even in the absence of the lattice torque, e.g., due to magnetic disorder at the interface.

In the absence of angular momentum transfer to the lattice, it is natural to assume axial symmetry with respect to magnetization direction which results in simplifications in Eqs.~(\ref{eq:Gv}), (\ref{eq:Gt1}), (\ref{eq:Gt2}), and (\ref{eq:Gt3}), i.e., $x_1^{\alpha(2)}=0$, $x_1^{\beta(0)}=x_2^{\beta(0)}$, $x_1^{\beta(1)}=x_2^{\beta(1)}=x_3^{\beta(1)}$, $x_2^{\beta(2)}=0$, $x_3^{\beta(2)}=x_4^{\beta(2)}=x_5^{\beta(2)}=x_1^{\beta(2)}$. This leads to the following generalization of Eq.~(\ref{eq:mixing}) for the spin mixing conductance:
\begin{equation}\label{eq:mixing1}
    \hat{{\cal G}}^{m}=2{\cal G}^{\uparrow\downarrow}_r(\hat{1}-\mathbf{m}\otimes\mathbf{m})+2{\cal G}^m_\parallel\mathbf{m}\otimes\mathbf{m}+2{\cal G}^{\uparrow\downarrow}_i \mathbf{m} \times,
\end{equation}
where ${\cal G}^{\uparrow\downarrow}_r= G_0\sum_{mn}\operatorname{Re}(\delta_{nm}-r_{mn}^{\uparrow\uparrow}r_{mn}^{\downarrow\downarrow*}-t_{mn}^{\uparrow\uparrow}t_{mn}^{\downarrow\downarrow*})$ describes the absorption of transverse spin current and ${\cal G}^m_\parallel=G_{0}(T_{\uparrow\downarrow}+T_{\downarrow\uparrow}+R_{\uparrow\downarrow}+R_{\downarrow\uparrow})$ the absorption of longitudinal spin current (i.e., spin memory loss); ${\cal G}^{\uparrow\downarrow}_i= G_0\sum_{mn}\operatorname{Im}(\delta_{nm}-r_{mn}^{\uparrow\uparrow}r_{mn}^{\downarrow\downarrow*}-t_{mn}^{\uparrow\uparrow}t_{mn}^{\downarrow\downarrow*})$ describes the precession of spins. Even though the formal expressions for ${\cal G}^{\uparrow\downarrow}_r$ and ${\cal G}^{\uparrow\downarrow}_i$ did not change compared to Eq.~(\ref{eq:mixing}), their values can still be affected by the presence of spin-flip scattering due to unitarity of the scattering matrix. The effect of the unitarity constraint, however, does not have a direct relation to the spin memory loss parameter $\delta$. \cite{Belashchenko.Kovalev.ea:PRL2016}

Using a typical spin-orbit torque geometry \cite{Chen.Takahashi.ea:2013} and Eq.\ \eqref{eq:spin}, we can write a boundary condition determining the torque:
\begin{equation}
    \frac{2 e^2}{\hbar}\vec{\tau}_F=e (\hat{1}-\mathbf{m}\otimes\mathbf{m}) \mathbf{j}^s=(\hat{1}-\mathbf{m}\otimes\mathbf{m})\hat{{\cal G}}^{m}\cdot \boldsymbol \mu^s,
\end{equation}
where $\boldsymbol \mu^s$ is the spin accumulation and $\vec{\tau}_F$ is the magnetization torque. The spin current can be further calculated from the diffusion equation:
\begin{equation}
    \nabla^2 \boldsymbol \mu^s=\boldsymbol \mu^s/l^2_{sf},
\end{equation}
and
\begin{equation}
   \mathbf{j}^s=- \frac{\sigma}{2e}\partial_z \boldsymbol \mu^s+j^{SH}\hat{y},
\end{equation}
where the interface is orthogonal to $z$ axis and $j^{SH}$ is the spin Hall current. We recover conventional antidamping and field like torques:
\begin{eqnarray}
   \vec{\tau}_F=(\hbar j^{SH}/2e)\left[ \frac{g^{\uparrow\downarrow}_r \tanh{\delta/2}}{1+2g^{\uparrow\downarrow}_r\coth{\delta}} \mathbf{m}\times(\mathbf{m}\times\hat{y}) \right.\\
   \left. +\frac{g^{\uparrow\downarrow}_i \tanh{\delta/2}}{1+2g^{\uparrow\downarrow}_i\coth{\delta}}\mathbf{m}\times\hat{y}\right],\nonumber
\end{eqnarray}
where $g^{\uparrow\downarrow}_{r(i)}=(l_{sf}/\sigma){\cal G}^{\uparrow\downarrow}_{r(i)}$ and $\sigma$ is the conductivity of the normal metal.
The results of this section are inconsistent with the notion that spin memory loss should directly affect spin-orbit torque. \cite{PhysRevLett.116.126601,PhysRevB.92.064426,PhysRevLett.112.106602,PhysRevLett.122.077201} As can be seen from Eq.\ \eqref{eq:mixing1}, two separate parameters are responsible for spin memory loss and spin-orbit torque, and in general there is no direct connection between the two. In the presence of spin-orbit interactions, only the total torque acting on the lattice and magnetization can be obtained from the circuit theory. However, it seems that a similar conclusion can be reached about the absence of direct relation between spin memory loss and torque.

\section{Computational details and interface geometry}
\label{methods}

The transmittances and reflectances (\ref{tss})-(\ref{rss}) were calculated using the Landauer-B\"uttiker approach implemented in the tight-binding linear muffin-tin orbital (LMTO) method. \cite{questaal} Spin-orbit coupling (SOC) was introduced as a perturbation to the LMTO potential parameters. \cite{Turek-SOC,questaal} Local density approximation (LDA) was used for exchange and correlation. \cite{Barth_1972}

We have considered a number of interfaces between metals with the face-centered cubic lattice. The interfaces were assumed to be epitaxial with the (111) or (001) crystallographic orientation. Lattice relaxations were neglected, and the average lattice parameter for the two lead metals was used for the given interface. The polarization of the spin current and the magnetization (in F|N and F|F systems) were taken to be either parallel or perpendicular to the interface.

Self-consistent charge and spin densities were obtained using periodic supercells with at least 12 monolayers of each metal. The surface Brillouin zone integration in transport calculations was performed with a $512\times512$ mesh for magnetic and $128\times128$ for non-magnetic systems.

We also studied the influence of interfacial intermixing on spin-memory loss at Pt$\mid$Pd and Au$\mid$Pd interfaces. One layer on each side of the interface was intermixed with the metal on the other side. The mixing concentrations were varied from 11\% to 50\%. For example, an A|B interface with 25\% intermixing had two disordered layers with compositions A$_{0.75}$B$_{0.25}$ and A$_{0.25}$B$_{0.75}$ between pure A and pure B leads. The transverse size of the supercell was $2\times2$ for 25\% and 50\% intermixing and $3\times3$ for 11\% intermixing. The conductances were averaged over all possible configurations in the $2\times2$ supercell and over 18 randomly generated configurations in $3\times3$. In addition, a model with long-range intermixing (LRI) was considered where the transition from pure A to pure B occurs over 8 intermixed monolayers with compositions A$_{8/9}$B$_{1/9}$, A$_{7/9}$B$_{2/9}$,\dots, A$_{1/9}$B$_{8/9}$. This model was implemented using 3$\times$3 supercells.

\section{Adiabatic embedding}

In the Landauer-B\"uttiker approach, the active region where scattering takes place is embedded between ideal semi-infinite leads. In the circuit theory, the leads are imagined to be built into the nodes of the circuit on both sides of the given interface. In order to define spin-dependent scattering matrices with respect to the well-defined spin bases, we turn off SOC in the leads.

To avoid spurious scattering at the boundaries with the SOC-free leads, we introduce ``ramp-up'' regions between the interface and the leads, wherein the SOC is gradually increased from zero at the edges of the active region to its actual magnitude near the interface. Specifically, for an atom at a distance $x$ from the interface ($|x|>l_0$), the SOC parameters are scaled by $(L-2|x|)/(L-2l_0)$, where $L$ is the total length of the active region and $l_0$ the length of the region on each side of the interface where SOC is retained at full strength. In our calculations we set $l_0$ to 2 monolayers.

Because a slowly varying potential only allows scattering with a correspondingly small momentum transfer, such \emph{adiabatic embedding} \cite{Belashchenko.Kovalev.ea:PRL2016} allows a generic pure spin state from the lead to evolve without scattering into the bulk eigenstate of the metal before being scattered at the interface.

In a non-magnetic metal, as explained in Ref.\ \onlinecite{Belashchenko.Kovalev.ea:PRL2016}, adiabatic embedding leads to strong reflection near the lines on the Fermi surface where the group velocity is parallel to the interface. Geometrically, when projected orthographically onto the plane of the interface, these lines form the boundaries of the projected Fermi surface. Electrons with such wave vectors can backscatter from the SOC ramp-up region both with and without a spin flip. The contribution of this backscattering to the spin-flip reflectance is an artefact of adiabatic embedding and needs to be subtracted out. \cite{Belashchenko.Kovalev.ea:PRL2016} In a magnetic lead such backscattering conserves spin and is, therefore, inconsequential for spin-memory loss calculations.

\begin{figure}[htb]
	\centering
	\includegraphics[width=0.9\columnwidth]{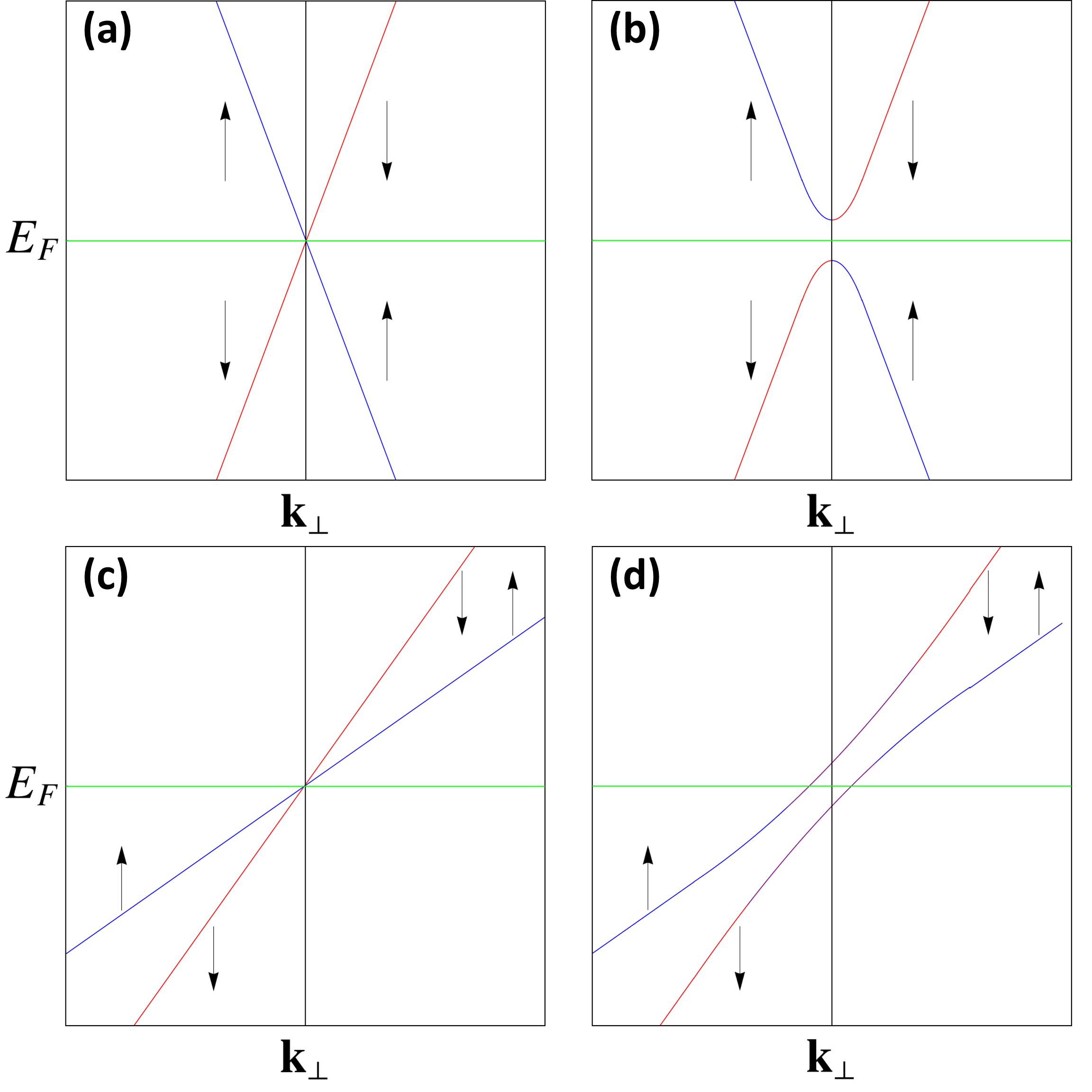}
	\caption{Crossing of the electronic bands in a ferromagnetic lead near an intersection of two Fermi surface sheets of opposite spin. The parallel component of the quasi-momentum, $\mathbf{k}_\parallel$, is fixed. (a-b) and (c-d): Cases where the normal component of the group velocity $v_\perp$ has the same or opposite sign on the two sheets, resulting in resonant spin-flip reflection or transmission, respectively. (a) and (c): no SOC; (b) and (d): avoided crossings induced by SOC.}
	\label{fig:soc}
\end{figure}

Adiabatic embedding can also produce strong scattering near the intersections of different sheets of the Fermi surface, where an electron can scatter from one sheet to another with a small momentum transfer. Such intersections do not exist in non-magnetic metals considered in this paper (Cu, Ag, Au, Pd, Pt), but they are present in all ferromagnetic transition metals. When the two intersecting sheets correspond to states of opposite spin, scattering from one sheet to the other is a spin-flip process. Depending on the signs of the normal (to the interface) components $v_\perp$ of the group velocities at the intersection, this scattering may or may not change the propagation direction with respect to the interface and thereby show up in spin-flip reflection or transmission. These two situations are illustrated in Fig.~\ref{fig:soc}. If $v_\perp$ has opposite signs on the two intersecting sheets [see Fig.~\ref{fig:soc}(a-b)], then SOC opens a gap at the avoided crossing, and incident electrons with quasi-momenta close to the intersection are fully reflected from the ramp-up region with a spin flip. On the other hand, if $v_\perp$ has the same sign on the two sheets [see Fig \ref{fig:soc}(c-d)], then, instead of backscattering, there is a large probability of forward spin-flip scattering as the electron passes through the ramp-up region.

Because we are interested in the spin-flip scattering processes introduced by the interface, the contribution of spin-flip scattering due to the presence of the ramp-up regions in the leads should be subtracted out. Unfortunately, this can only be done approximately. The approach used for N$_1$|N$_2$ interfaces in Ref.\ \onlinecite{Belashchenko.Kovalev.ea:PRL2016} was to subtract the spin-flip reflectances of auxiliary systems N$_1$|N$_1$ and N$_2$|N$_2$ where the same lead material is used on both sides of an imaginary interface with adiabatic embedding. This method is reasonable because the electrons incident from one of the leads and backscattered by the ramp-up region never reach the interface in the real N$_1$|N$_2$ system. In an F|N system, the same is true for the backscattering on Fermi sheet crossings in F [the case of Fig.~\ref{fig:soc}(a-b)], but not for the forward scattering [the case of Fig.~\ref{fig:soc}(c-d)].

Nevertheless, as a simple approximation, we extend the approach of Ref.\ \onlinecite{Belashchenko.Kovalev.ea:PRL2016} to the F|N interfaces, subtracting both the spin-flip reflectances in auxiliary F|F and N|N systems and the spin-flip transmittance in auxiliary F|F. Likewise, for an F$_1$|F$_2$ interface, we subtract both reflectances and transmittances in F$_1$|F$_1$ and F$_2$|F$_2$. Thus, for any kind of interface, we define
\begin{align}
  &  T'_{\uparrow\downarrow}=T^{1|2}_{\uparrow\downarrow}-T^{1|1}_{\uparrow\downarrow}-T^{2|2}_{\uparrow\downarrow}\\
  &  R'_{a,\uparrow\downarrow}=R^{1|2}_{a,\uparrow\downarrow}-R^{a|a}_{\uparrow\downarrow},
\end{align}
where $a=L$ or $a=R$ denotes one of the leads, and the primed quantities are used in Eq.\ (\ref{eq:approx-1}).
In the following, we refer to this as the subtraction method, and the parameter $\delta$ calculated in this way is denoted $\delta_s$.

\subsection{$k$-point filtering}

\begin{figure*}[htb]
	\centering
    \includegraphics[width=0.9\textwidth]{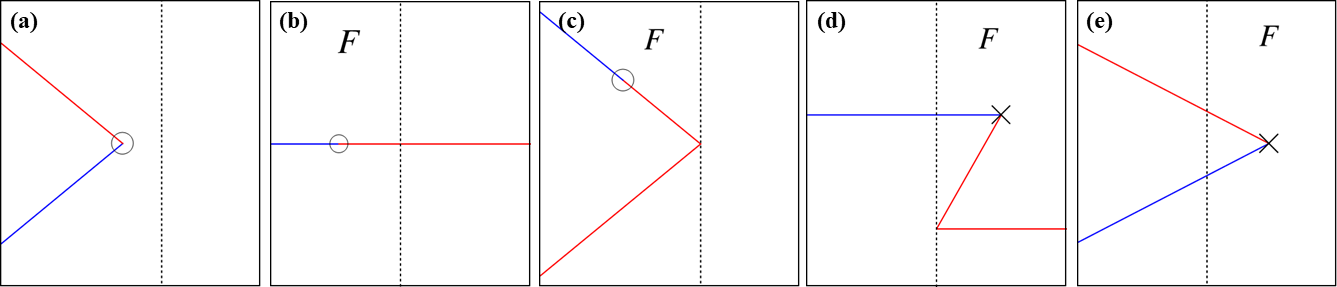}
	\caption{Spin-flip scattering mechanisms induced by a crossing of two Fermi sheets of opposite spin in an adiabatically embedded interface with no disorder. Dashed vertical lines show the interface; the label $F$ specifies that the given metal must be ferromagnetic. Blue and red lines schematically show the trajectory of an electron before and after the spin flip. Crosses show physical spin-flip scattering processes, while circles denote those that occurs solely due to adiabatic embedding.}
	\label{SF_mech}
\end{figure*}

A more fine-grained approach is to identify the locations in the surface Brillouin zone where spurious reflection or transmission occurs and filter out the contributions to spin-flip scattering probabilities from those locations. This filtering requires care, because some spin-flip scattering processes near the Fermi surface crossings are, in fact, physical, rather than merely being artefacts of adiabatic embedding.
This can be seen from Fig.\ \ref{SF_mech}, which shows possible spin-flip scattering processes facilitated by the crossing of the Fermi sheets of opposite spin.

Figure \ref{SF_mech}(a) shows a spin-flip backscattering process in the left lead, which can occur near a Fermi projection boundary in a normal metal or near a Fermi crossing of the type shown in Fig.\ \ref{fig:soc}(b). The processes shown in Figs. \ref{SF_mech}(b) and \ref{SF_mech}(c) result from the forward scattering near a Fermi crossing of the type shown in Fig.\ \ref{fig:soc}(d) in the left lead, where the electron is then either transmitted through or reflected from the interface, respectively. Each process has a reciprocal version. The three processes shown in Figs.\ \ref{SF_mech}(a-c) exist solely due to the presence of a ramp-up region, which provides the small momentum transfer needed to scatter from one Fermi sheet to another.

In contrast, Figs.\ \ref{SF_mech}(d) and \ref{SF_mech}(e) show physical scattering processes. Here, the momentum of an electron incident from the left lead lies inside the spin-orbit gap of the type shown in Fig.\ \ref{fig:soc}(b) in the right lead. As a result, the electron experiences a resonant spin-flip transmission [Fig.\ \ref{SF_mech}(d)] or reflection [Fig.\ \ref{SF_mech}(e)] at the interface. Resonant spin-flip transmission shown in Fig.\ \ref{SF_mech}(d) is possible because an electron can scatter to a different Fermi sheet with a large momentum transfer acquired from the interface. Illustrations in Fig.\ \ref{SF_mech}(d-e) are highly schematic because the wavefunction inside the spin-orbit gap is evanescent in the right lead.

\begin{figure*}[htb]
	\centering
	\includegraphics[width=0.9\textwidth]{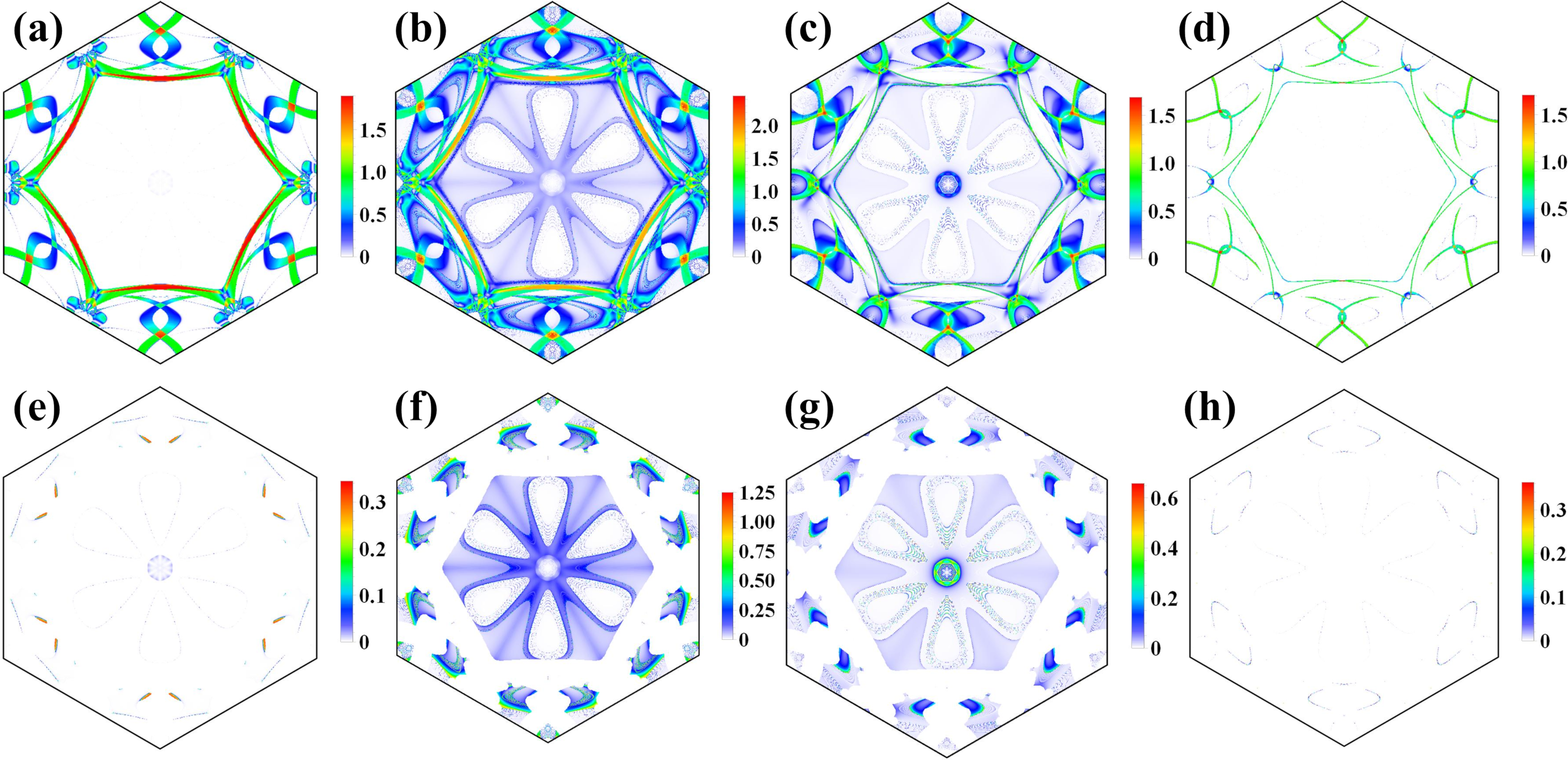}
	\caption{$k$-resolved spin-flip reflection functions for adiabatically embedded Pt|Pt, Pd|Pd, and Pt|Pd interfaces with and without $k$-point filtering. (a) $R_{\downarrow\uparrow}$ in Pt$\mid$Pt; (b) $R_{L\downarrow\uparrow}$ in Pt$\mid$Pd; (c) $R_{R \downarrow\uparrow}$ in Pt$\mid$Pd; (d) $R_{\downarrow\uparrow}$ in Pd$\mid$Pd; (e) $R_{\downarrow\uparrow}$ in Pt$\mid$Pt, filtered; (f) $R_{L \downarrow\uparrow}$ in Pt$\mid$Pd, filtered; (g) $R_{R \downarrow\uparrow}$ in Pt$\mid$Pd, filtered; (h)  $R_{\downarrow\uparrow}$ in Pd$\mid$Pd, filtered.}
	\label{fig:ptpd}
\end{figure*}

\begin{figure*}[htb]
	\centering
	\includegraphics[width=0.9\textwidth]{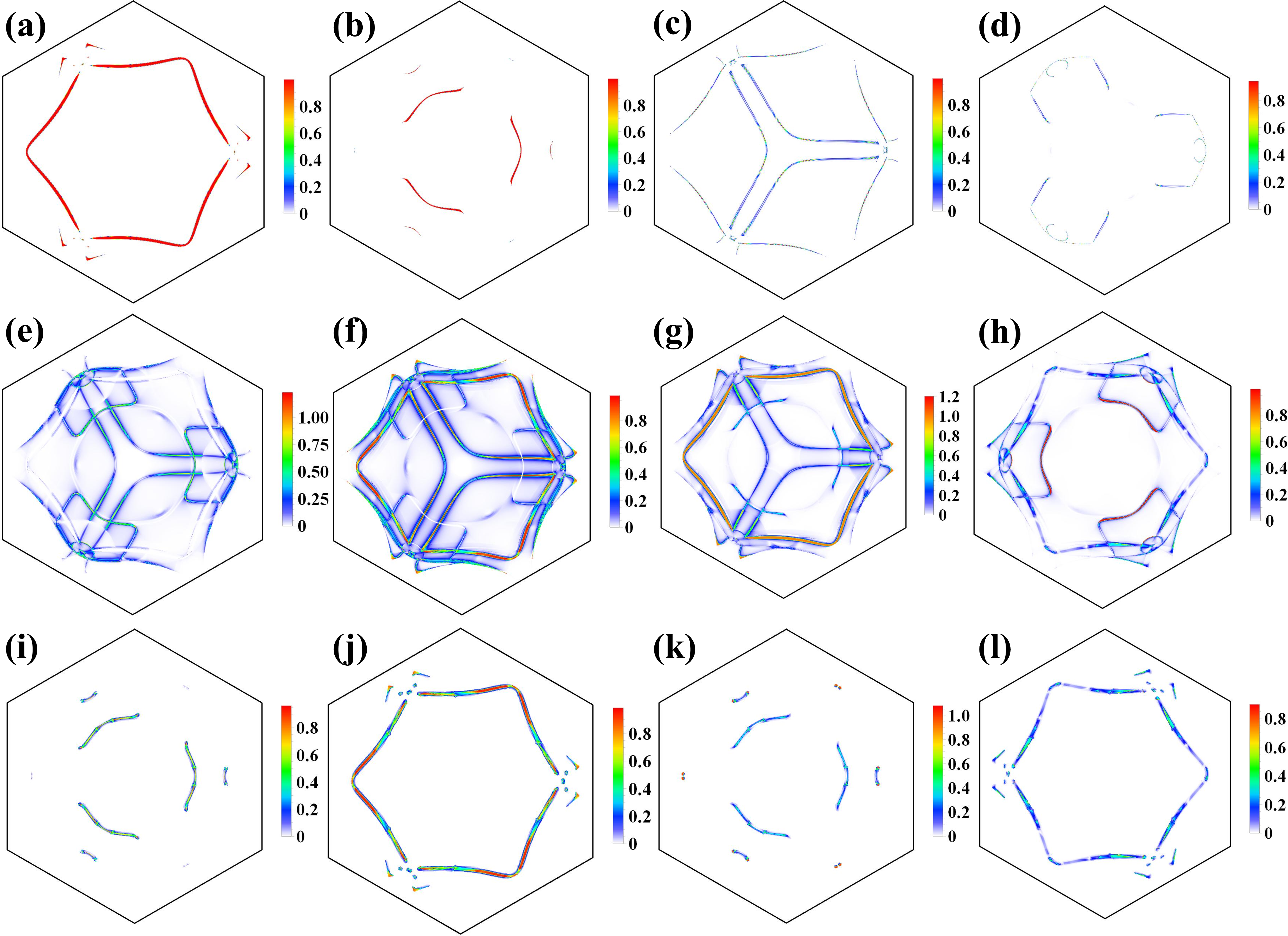}
	\caption{$k$-resolved spin-flip transmission and reflection functions for Ni|Ni, Co|Co, and Ni|Co, and an illustration of $k$-point filtering. (a) $R_{\downarrow\uparrow}$ in Ni$\mid$Ni; (b) $R_{\downarrow\uparrow}$ in Co$\mid$Co; (c) $T_{\downarrow\uparrow}$ in Ni$\mid$Ni; (d) $T_{\downarrow\uparrow}$ in Co|Co; (e) $T_{\downarrow\uparrow}$ in Ni|Co; (f) $T_{\uparrow\downarrow}$ in Ni|Co; (g) $R^{L}_{\downarrow\uparrow}$ in Ni$\mid$Co; (h) $R^{R}_{\downarrow\uparrow}$ in Ni$\mid$Co; (i) $T_{\downarrow\uparrow}$ in Ni$\mid$Co, filtered; (j) $T_{\uparrow\downarrow}$ in Ni$\mid$Co, filtered; (k) $R^{L}_{\downarrow\uparrow}$ in Ni$\mid$Co, filtered; (l) $R^{R}_{\downarrow\uparrow}$ in Ni$\mid$Co, filtered.}
	\label{CONI}
\end{figure*}

Let us first examine the spin-flip scattering processes in systems without a physical interface, where all scattering is due to adiabatic embedding alone. Spin-flip reflection at the Fermi projection boundaries can be seen in Figs.\ \ref{fig:ptpd}(a) and \ref{fig:ptpd}(d) for adiabatically embedded Pt and Pd, respectively, denoted in the figure caption as a fictitious ``interface'' of a material with itself (e.g., Pd|Pd). \cite{Belashchenko.Kovalev.ea:PRL2016} The areas with strong spin-flip reflection are notably broader in Pt, which has a larger spin-orbit constant compared to Pd. Spin-flip reflection at Fermi crossings can be seen in Figs.\ \ref{CONI}(a) and \ref{CONI}(b) for adiabatically embedded Ni and Co, respectively. These two cases correspond to the diagram in Fig.\ \ref{SF_mech}(a). Spin-flip transmission at Fermi crossings in Ni and Co is seen, in turn, in Figs.\ \ref{CONI}(c) and \ref{CONI}(d); this is the process shown in Fig.\ \ref{SF_mech}(b) without the physical interface.

Now consider physical interfaces. Contours with strong spin-flip reflection in, say, Fig.\ \ref{fig:ptpd}(d) for Pd|Pd are also seen in Fig.\ \ref{fig:ptpd}(c) for electrons incident from the Pd lead in Pt|Pd; the same comparison can be made for contours with strong spin-flip reflection in, say, Fig.\ \ref{CONI}(a) for Ni|Ni and \ref{CONI}(g) for Ni|Co. These processes correspond to Fig.\ \ref{fig:soc}(a). Furthermore, the contours with strong spin-flip transmission in Fig.\ \ref{CONI}(c) for Ni|Ni show up in both Fig.\ \ref{CONI}(e) and \ref{CONI}(g) for spin-flip transmission and reflection in Ni|Co, respectively. These processes correspond to Fig.\ \ref{SF_mech}(b) and \ref{SF_mech}(c). The contours with resonant spin-flip transmission in Co|Co [Fig.\ \ref{CONI}(d)] also show up in spin-flip transmission for Ni|Co [Fig.\ \ref{CONI}(e)]; this corresponds to Fig.\ \ref{SF_mech}(b) with the two leads interchanged.

All of the spin-flip scattering processes mentioned so far and corresponding to Fig.\ \ref{SF_mech}(a-c) are artefacts of adiabatic embedding and need to be filtered out in the calculation of the interfacial spin loss parameter. On the other hand, the spin-flip transmission [Fig.\ \ref{CONI}(e)] and reflection [Fig.\ \ref{CONI}(g)] functions for the Ni|Co interface also show the spin-flip resonances of the types shown in Fig.\ \ref{SF_mech}(d-e). Consider the spin-flip reflection function for electrons incident from the Ni lead for the Ni|Co interface, which is shown in Fig.\ \ref{CONI}(g). Apart from the resonant contours appearing in Fig.\ \ref{CONI}(a) and \ref{CONI}(c) for spin-flip reflection and transmission in Ni|Ni, there are also resonant contours in Fig.\ \ref{CONI}(g) that correspond to the spin-flip reflection resonances in Co|Co, which are seen in Fig.\ \ref{CONI}(b). The same resonant contours appearing in Fig.\ \ref{CONI}(e) for the spin-flip transmission in Ni|Co correspond to the process shown in Fig.\ \ref{SF_mech}(d). These resonances correspond to the \emph{physical} process depicted in Fig.\ \ref{SF_mech}(e) and should \emph{not} be filtered out in the calculation of the spin loss parameter.

This analysis shows that both artefacts of adiabatic embedding [Fig.\ \ref{SF_mech}(a-c)] and physical resonant spin-flip scattering processes [Fig.\ \ref{SF_mech}(d-e)] can be located in $k$-space using spin-flip transmission functions calculated for auxiliary systems. Thus, as an alternative to the subtraction method discussed above, the artefacts of adiabatic embedding can be removed using $k$-point filtering.

For nonmagnetic (N$_1$|N$_2$) interfaces, we first identify the $k$-points where the spin-flip reflectance in an auxiliary system (N$_1$|N$_1$ or N$_2$|N$_2$) exceeds a certain threshold value, which is chosen so that the spin-flip reflectance in the auxiliary system becomes less than $0.001G_0$ if the contributions from the identified $k$-points are excluded. Then the contributions from those $k$-points are excluded in the calculation of the spin-flip reflectance for electrons incident from the corresponding lead. To ensure that the artefacts are fully removed, the excluded regions are slightly enlarged.

Ferromagnetic leads induce resonant scattering near the crossings of the Fermi surfaces for opposite spins. Processes of the types shown in Fig. \ref{SF_mech}(a-c) should be filtered out, as explained above. We found that the spin-flip reflectances and transmittances for all ferromagnetic interfaces considered here are dominated by resonant processes depicted in Fig.\ \ref{SF_mech}(d-e) rather than by contributions from generic $k$-points. Indeed, the spin-loss parameters obtained by excluding the processes of Fig.\ \ref{SF_mech}(a-c) or by including only those in Fig.\ \ref{SF_mech}(d-e) are almost identical.
Figures \ref{CONI}(i-l) show the spin-flip scattering functions obtained by starting from Figs.\ \ref{CONI}(e-h) and filtering out everything other than the processes of Fig.\ \ref{SF_mech}(d-e). By performing $k$-point filtering in this way we obtain a lower bound on the spin-flip scattering functions and the spin-loss parameter, ensuring that the artefacts of adiabatic embedding are completely removed. The values $\delta_{f}$ listed in Table \ref{tab:3} were obtained in this way.

\section{Results}

\subsection{Non-magnetic interfaces}

Table \ref{tab:1} lists the area-resistance products $AR$ and the spin-loss parameters for nonmagnetic interfaces. The subtraction and $k$-point filtering methods result in similar values of $\delta$. For all material combinations, $\delta$ is quite similar for (001) and (111) interfaces, suggesting that the crystallographic structure of the interface does not have a strong effect on interfacial spin relaxation. In all cases, the spin-loss parameter is slightly lower for the parallel orientation of the spin accumulation relative to the interface.

The calculated $AR$ products and $\delta$ parameters are in good agreement with experimental measurements \cite{Bass:JMMM2016} in systems without Pd, but both are strongly overestimated for (Au,Ag,Cu,Pd)|Pd interfaces. However, the results for the Au|Pd (111) interface with the spin accumulation parallel to the interface are in good agreement with recent calculations of Gupta \emph{et al.} \cite{Gupta2020} ($AR=0.81$ f$\Omega\cdot$m$^2$ and $\delta=0.43$) based on the analysis of the local spin currents near the interface.

\begin{table}[ht]
    \caption{Area-resistance products $AR$ (f$\Omega\cdot$m$^2$) and spin-loss parameters obtained using the subtraction method ($\delta_s$) and the filtering method ($\delta_f$) for nonmagnetic interfaces. $\mathbf{M}$ denotes the orientation of the spin accumulation relative to the interface.}
	\centering
	\begin{tabular}{|c|c|c|c|c|c|c|c|c|}
		\hline
		N$\mid$N & Plane & $\textbf{M}$ & $AR$ & $AR_{exp}$ & $\delta_{s}$ & $\delta_{f}$ & $\delta_{exp}$ \\
		\hline
		\multirow{4}{*}{Pt$\mid$Pd} & \multirow{2}{*}{001} & $\parallel$ &  0.42 & \multirow{4}{*}{0.14$\pm$0.03} & 0.60 & 0.57 & \multirow{4}{*}{0.13$\pm$0.08}\\
		\cline{3-4} \cline{6-7}
		& & $\bot$ & 0.44 &  & 0.71 & 0.65 & \\
		\cline{2-4} \cline{6-7}
		& \multirow{2}{*}{111} & $\parallel$ &0.28 &  & 0.41 & 0.36 &  \\
		\cline{3-4} \cline{6-7}
		& & $\bot$ & 0.29 &  & 0.45 & 0.38 & \\
		\hline
		\multirow{4}{*}{Au$\mid$Pd} & \multirow{2}{*}{001} & $\parallel$ &  0.96 & \multirow{4}{*}{0.23$\pm$0.08} & 0.71 &0.68 & \multirow{4}{*}{0.08$\pm$0.08}\\
		\cline{3-4} \cline{6-7}
		& & $\bot$ & 0.96 &  & 0.86 & 0.82 & \\
		\cline{2-4} \cline{6-7}
		& \multirow{2}{*}{111} & $\parallel$ &0.83 &  & 0.53 & 0.54 &  \\
		\cline{3-4} \cline{6-7}
		& & $\bot$ & 0.87 &  & 0.73 & 0.69 & \\
		\hline
		\multirow{4}{*}{Ag$\mid$Pd} & \multirow{2}{*}{001} & $\parallel$ &  0.92 & \multirow{4}{*}{0.35$\pm$0.08} & 0.41 & 0.47 & \multirow{4}{*}{0.15$\pm$0.08}\\
		\cline{3-4} \cline{6-7}
		& & $\bot$ & 1.12 &  & 0.50 & 0.54 & \\
		\cline{2-4} \cline{6-7}
		& \multirow{2}{*}{111} & $\parallel$ &0.89 &  & 0.41 & 0.47 &  \\
		\cline{3-4} \cline{6-7}
		& & $\bot$ & 0.92 &  & 0.50 & 0.55 & \\
		\hline
		\multirow{4}{*}{Cu$\mid$Pd} & \multirow{2}{*}{001} & $\parallel$ &  0.81 & \multirow{4}{*}{0.45$\pm$0.005} & 0.41 & 0.47 & \multirow{4}{*}{0.24$\pm$0.05}\\
		\cline{3-4} \cline{6-7}
		& & $\bot$ & 0.81 &  & 0.47 & 0.52 & \\
		\cline{2-4} \cline{6-7}
		& \multirow{2}{*}{111} & $\parallel$ &0.80 &  & 0.43 & 0.40 &  \\
		\cline{3-4} \cline{6-7}
		& & $\bot$ & 0.81 &  & 0.53 & 0.48 & \\
		\hline
		\multirow{4}{*}{Cu$\mid$Au} & \multirow{2}{*}{001} & $\parallel$ &  0.13 & \multirow{4}{*}{0.15$\pm$0.005} & 0.08 & 0.08 & \multirow{4}{*}{0.13$\pm$0.07}\\
		\cline{3-4} \cline{6-7}
		& & $\bot$ & 0.13 &  & 0.11 & 0.11 & \\
		\cline{2-4} \cline{6-7}
		& \multirow{2}{*}{111} & $\parallel$ &0.11 &  & 0.08 & 0.07 &  \\
		\cline{3-4} \cline{6-7}
		& & $\bot$ & 0.12 &  & 0.11 & 0.10 & \\
		\hline
		\multirow{4}{*}{Cu$\mid$Pt} & \multirow{2}{*}{001} & $\parallel$ &  0.90 & \multirow{4}{*}{0.75$\pm$0.05} & 1.00 & 0.87 & \multirow{4}{*}{0.9$\pm$0.1}\\
		\cline{3-4} \cline{6-7}
		& & $\bot$ & 0.89 &  & 1.07 & 0.9 & \\
		\cline{2-4} \cline{6-7}
		& \multirow{2}{*}{111} & $\parallel$ &0.75 &  & 0.88 & 0.72 &  \\
		\cline{3-4} \cline{6-7}
		& & $\bot$ & 0.82 &  & 1.11 & 0.83 & \\
		\hline
		\multirow{4}{*}{Cu$\mid$Ag} & \multirow{2}{*}{001} & $\parallel$ &  0.03 & \multirow{4}{*}{0.045$\pm$0.005} & 0.02 & 0.2 & \multirow{4}{*}{0}\\
		\cline{3-4} \cline{6-7}
		& & $\bot$ & 0.03 &  & 0.03 & 0.02 & \\
		\cline{2-4} \cline{6-7}
		& \multirow{2}{*}{111} & $\parallel$ &0.13 &  & 0.03 & 0.03 &  \\
		\cline{3-4} \cline{6-7}
		& & $\bot$ & 0.13 &  & 0.04 & 0.04 & \\
		\hline
	\end{tabular}
	\label{tab:1}
\end{table}

The large discrepancy in $AR$ for interfaces with Pd suggests that the idealized interface model is inadequate for these interfaces. Therefore, Pt$\mid$Pd and Au$\mid$Pd with interfacial intermixing were also constructed as described in Section \ref{methods}. The results for intermixed interfaces are listed in Table \ref{tab:2}. It is notable that intermixing increases the $AR$ product, while its values for ideal interfaces with Pd are already too large compared with experimental reports. The spin-loss parameter $\delta$ is also significantly increased by intermixing, which moves it further away from experimental data.

\begin{table}[ht]
    \caption{Same as in Table \ref{tab:1} but for non-magnetic interfaces with intermixing. The percentage indicates the composition in the two intermixed layers. LRI refers to the long-range intermixing model; see Section \ref{methods} for details.}
	\centering
	\begin{tabular}{|c|c|c|c|c|c|c|c|c|}
		\hline
		N$\mid$N (mix \%)& Plane & $\textbf{M}$ & AR & $AR_{exp}$ & $\delta_{s}$ & $\delta_{f}$ & $\delta_{exp}$ \\
		\hline
		\multirow{2}{*}{Pt$\mid$Pd (11\%)} & \multirow{2}{*}{111} & $\parallel$ &  0.29 & \multirow{8}{*}{0.14$\pm$ 0.03} & 0.45 & 0.38 & \multirow{8}{*}{0.13$\pm$0.08}\\
		\cline{3-4} \cline{6-7}
		& & $\bot$ & 0.30 &  & 0.56 & 0.40 & \\
		\cline{1-4} \cline{6-7}
		\multirow{2}{*}{Pt$\mid$Pd (25\%)} & \multirow{2}{*}{111} & $\parallel$ &  0.32 &  & 0.52 & 0.46 & \\
		\cline{3-4} \cline{6-7}
		& & $\bot$ & 0.34 &  & 0.65 & 0.52 & \\
		\cline{1-4} \cline{6-7}
		\multirow{2}{*}{Pt$\mid$Pd (50\%)} & \multirow{2}{*}{111} & $\parallel$ &  0.36 &  & 0.58 & 0.51 & \\
		\cline{3-4} \cline{6-7}
		& & $\bot$ & 0.38 &  & 0.72 & 0.57 & \\
		\cline{1-4} \cline{6-7}
		\multirow{2}{*}{Pt$\mid$Pd (LRI)} & \multirow{2}{*}{111} & $\parallel$ &  0.82 &  & 1.20 & 0.91 & \\
		\cline{3-4} \cline{6-7}
		& & $\bot$ & 0.85 &  & 1.34 & 0.96 & \\
		\hline
		\multirow{2}{*}{Au$\mid$Pd (11\%)} & \multirow{2}{*}{111} & $\parallel$ &  0.86 & \multirow{8}{*}{0.23$\pm$0.08} & 0.56 & 0.46 & \multirow{8}{*}{0.08$\pm$0.08}\\
		\cline{3-4} \cline{6-7}
		& & $\bot$ & 0.90 &  & 0.76 & 0.58 & \\
		\cline{1-4} \cline{6-7}
		\multirow{2}{*}{Au$\mid$Pd (25\%)} & \multirow{2}{*}{111} & $\parallel$ &  0.96 &  & 0.60 & 0.58 & \\
		\cline{3-4} \cline{6-7}
		& & $\bot$ & 1.01 &  & 0.81 & 0.73 & \\
		\cline{1-4} \cline{6-7}
		\multirow{2}{*}{Au$\mid$Pd (50\%)} & \multirow{2}{*}{111} & $\parallel$ &  0.95 &  & 0.60 & 0.58 & \\
		\cline{3-4} \cline{6-7}
		& & $\bot$ & 0.99 &  & 0.82 & 0.73 & \\
		\cline{1-4} \cline{6-7}
		\multirow{2}{*}{Au$\mid$Pd (LRI)} & \multirow{2}{*}{111} & $\parallel$ &  1.24 &  & 0.79 & 0.65 & \\
		\cline{3-4} \cline{6-7}
		& & $\bot$ & 1.29 &  & 0.98 & 0.76 & \\
		\hline
	\end{tabular}
	\label{tab:2}
\end{table}

The disagreement with experiment in the values of $AR$ and $\delta$ for interfaces with Pd is likely due to the lack of understanding of the interfacial structure in the sputtered multilayers, for which no structural characterization is available, to out knowledge. It seems somewhat implausible that the real sputtered interfaces are much less resistive compared to both ideal or intermixed interfaces considered here. It is possible that nominally bulk regions in sputtered multilayers containing Pd are more disordered and thereby have a higher resistivity and shorter spin-diffusion length compared to pure Pd films. The fitting procedure used to extract the $AR$ and $\delta$ parameters for the interface \cite{Bass:JMMM2016} would then ascribe this additional bulk resistance and spin relaxation to the interfaces.

\subsection{Ferromagnetic interfaces}

Table \ref{tab:3} lists the results for interfaces with one or two ferromagnetic leads. The $AR$ products for all interfaces are in excellent agreement with experimental data. \cite{Bass:JMMM2016} The values of the spin-loss parameter obtained using the subtraction method ($\delta_s$) tend to be larger, by up to a factor of 2, compared to the $k$-point filtering method ($\delta_f$), which is expected to be more accurate. For Pt|Co the results for $AR$ and $\delta$ are in good agreement both with experiment and with calculations using the discontinuity of the spin current. \cite{Gupta2020} In other systems $AR$ agrees very well with experiment but $\delta$ is underestimated, which may be due to the neglect of interfacial disorder and to the limitations of the adiabatic embedding method.

\begin{table}[ht]
	\caption{Same as in Table \ref{tab:1} but for F|N and F|F interfaces.}
	\centering
	\begin{tabular}{|c|c|c|c|c|c|c|c|c|c|c|}
		\hline
		F(N)$\mid$F & Plane & $\textbf{M}$ & $AR_{\uparrow}$ & $AR_{\downarrow}$ & AR & $AR_{exp}$ & $\delta_{s}$ & $\delta_{f}$ & $\delta_{exp}$ \\
		\hline
		\multirow{4}{*}{Cu$\mid$Co} & \multirow{2}{*}{001} & $\parallel$ &  0.29 & 2.06 & 0.59 & \multirow{4}{*}{\shortstack[r]{0.51\\$\pm$0.05}} & 0.22 & 0.12 & \multirow{4}{*}{\shortstack[r]{0.33 \\ $\pm$0.05}}\\
		\cline{3-6} \cline{8-9}
		& & $\bot$ & 0.31 & 2.05 & 0.59 &  & 0.24 & 0.14 & \\
		\cline{2-6} \cline{8-9}
		& \multirow{2}{*}{111} & $\parallel$ & 0.36 & 1.54 & 0.48 &  & 0.18 & 0.11 &  \\
		\cline{3-6} \cline{8-9}
		& & $\bot$ & 0.36 & 1.52 & 0.47 &  & 0.19 & 0.12 & \\
		\hline
		\multirow{4}{*}{Pt$\mid$Co} & \multirow{2}{*}{001} & $\parallel$ &  0.46 & 4.67 & 1.28  & \multirow{4}{*}{\shortstack[r]{0.85\\$\pm$0.12}} & 1.12 & 0.91 & \multirow{4}{*}{\shortstack[r]{0.9 \\ $\pm$0.4}}\\
		\cline{3-6} \cline{8-9}
		& & $\bot$ & 0.44 & 4.60 & 1.26 &  & 1.17 & 0.96 & \\
		\cline{2-6} \cline{8-9}
		& \multirow{2}{*}{111} & $\parallel$ & 1.70 & 1.36 & 0.76 &  & 0.81 & 0.72 &  \\
		\cline{3-6} \cline{8-9}
		& & $\bot$ & 1.82 & 1.38 & 0.80 &  & 0.91 & 0.80 & \\
		\hline
		\multirow{4}{*}{Ag$\mid$Co} & \multirow{2}{*}{001} & $\parallel$ &  0.40 & 1.87 & 0.57  & \multirow{4}{*}{\shortstack[r]{0.56 \\ $\pm$0.06}} & 0.33 & 0.21 & \multirow{4}{*}{\shortstack[r]{0.33 \\ $\pm$0.1}}\\
		\cline{3-6} \cline{8-9}
		& & $\bot$ & 0.43 & 1.84 & 0.57 &  & 0.38 & 0.29 & \\
		\cline{2-6} \cline{8-9}
		& \multirow{2}{*}{111} & $\parallel$ & 0.22 & 1.58 & 0.45 &  & 0.20 & 0.12 &  \\
		\cline{3-6} \cline{8-9}
		& & $\bot$ & 0.22 & 1.57 & 0.45 &  & 0.21 & 0.13 & \\
		\hline
		\multirow{4}{*}{Ni$\mid$Co} & \multirow{2}{*}{001} & $\parallel$ &  0.22 & 1.04 & 0.32  & \multirow{4}{*}{\shortstack[r]{0.255 \\ $\pm$0.025}} & 0.32 & 0.15 & \multirow{4}{*}{\shortstack[r]{0.35 \\ $\pm$0.05}}\\
		\cline{3-6} \cline{8-9}
		& & $\bot$ & 0.24 & 1.02 & 0.32 &  & 0.34 & 0.16 & \\
		\cline{2-6} \cline{8-9}
		& \multirow{2}{*}{111} & $\parallel$ & 0.21 & 0.73 & 0.23 &  & 0.27 & 0.17 &  \\
		\cline{3-6} \cline{8-9}
		& & $\bot$ & 0.25 & 0.72 & 0.24 &  & 0.29 & 0.16 & \\
		\hline
	\end{tabular}
	\label{tab:3}
\end{table}

\section{Conclusions}

We have developed a general formalism for analyzing magnetoelectronic circuits with spin-nonconserving N|N, F|N, or F|F interfaces between diffusive bulk regions. A tensor generalization of the spin mixing conductance encodes all possible spin-nonconserving processes, such as spin dephasing, spin loss, and spin precession. In the special case when exchange interactions dominate, those contributions can be clearly separated into terms responsible for spin memory loss, spin-orbit torque, and spin precession. Surprisingly, there is no direct relation between spin-orbit torque and spin memory loss; the two effects are described by different combinations of scattering amplitudes responsible for the absorption of the transverse and longitudinal components of spin current at the interface.

The spin relaxation (i.e., spin memory loss) parameter $\delta$ has been numerically calculated using Eqs.~(\ref{eq:approx}) and (\ref{eq:approx-1}) for a number of N|N, F|N, and F|F interfaces. First-principles calculations, aided by adiabatic embedding, show reasonable agreement with experiment for $\delta$ and the area-resistance products with the exception of N|N interfaces including a Pd lead. For such interfaces both $\delta$ and $AR$ are strongly overestimated, which can not be explained by short or long-range interfacial intermixing. The analysis of spin-flip scattering probabilities for F|N and F|F interfaces suggests that interfacial spin relaxation is dominated by electronic states near the crossings of the Fermi surfaces for opposite spins in ferromagnets.

The generalized magnetoelectronic circuit theory provides a convenient framework for analyzing spin transport in magnetic nanostructures with strong spin-orbit coupling at interfaces.

\begin{acknowledgments}
A. K. is much indebted to Gerrit Bauer for stimulating discussions on circuit theory with spin-flip scattering. This work was supported by the National Science Foundation through Grant No. DMR-1609776 and the Nebraska MRSEC, Grant No. DMR-1420645, as well as by the DOE Early Career Award DE-SC0014189 (AK) and the EPSRC CCP9 Flagship project, EP/M011631/1 (MvS). Computations were performed utilizing the Holland Computing Center of the University of Nebraska, which receives support from the Nebraska Research Initiative.
\end{acknowledgments}

\bibliographystyle{apsrev}
\bibliography{spin-loss}

\end{document}